\documentclass[a4paper,10pt]{article}
\usepackage[position=t,singlelinecheck=off]{subfig}
\usepackage{amssymb,amsmath,graphics,graphicx,float,braket}
\usepackage[colorlinks=true, citecolor=blue, urlcolor=blue ]{hyperref}
\usepackage{cite}
\def\beq{\begin{equation}}
\def\eeq{\end{equation}}
\def\bea{\begin{eqnarray}}
\def\eea{\end{eqnarray}}
\def\nn{\nonumber}

\makeatletter
\def\@cite#1#2{${\mbox{#1\if@tempswa , #2\fi}}$}
\makeatother

\begin{document}
\thispagestyle{empty}
\begin{center}
\begin{LARGE}
\textsf{Enhancement of squeezing in the Rabi model with parametric nonlinearity}
\end{LARGE} \\

\bigskip\bigskip
V. Yogesh$^{\dagger}$ and Prosenjit Maity$^{*}$ 
\\
\begin{small}
\bigskip
\textit{
 $^{\dagger}$S. N. Bose National Centre for Basic Sciences, \\ Block-JD, Sector-III, Salt Lake, Kolkata 700106, India. \\
$^{*}$ Department of Physics, Ramakrishna Mission Residential College, \\ Narendrapur, Kolkata-700103, India.}
\end{small}
\end{center}

\vfill
\begin{abstract}
The squeezing effect arises in the interacting qubit-oscillator system is studied with the presence of a parametric oscillator in the Rabi model. Based on the generalized rotating wave approximation which works well in the wide range of coupling strength as well as detuning, the analytically derived approximate energy spectrum is compared with the numerically determined spectrum of the Hamiltonian. For the initial state of the bipartite system, the dynamical evolution of the reduced density matrix corresponding to the oscillator is obtained by partial tracing over the qubit degree of freedom. The oscillator's reduced density matrix yields the nonnegative phase space quasi probability distribution known as Husimi $Q$-function which is utilized to compute the quadrature variance. It is shown that the squeezing produced in the Rabi model can be enhanced substantially in the presence of a parametric nonlinear term.

\end{abstract} 
 
\newpage
\setcounter{page}{1}

\section{Introduction}

The phenomenon of squeezing which is one of the signature of nonclassicality [\cite{Mandel1982}] was extensively studied in the past decades. Due to the attractive feature that the quantum fluctuations in one quadrature component of the field can be reduced below the standard quantum limit, the squeezed states of light provide potential applications including high-precision quantum measurements [\cite{CM1981},\cite{MA2013}], quantum communication [\cite{SLB2005}], enhanced sensitivity in gravitational wave detectors [\cite{JAS2013}] etc. In the quantum optical domain, squeezed light has been more commonly generated using nonlinear optical processes, including degenerate parametric amplification and degenerate four-wave mixing [\cite{DFW2007,HP1978,RES1985,RMS1986,LAW1986}]. The authors of [\cite{PMZ1982}] showed that squeezing of a single mode quantized electromagnetic field could be obtained in the Jaynes-Cummings model [\cite{JC1963}] of a resonant two-level atom interacting with the field prepared initially in the coherent state. Subsequently it has been found that the significant amount of squeezing in the Jaynes-Cummings model can happen only when the mean photon number in the field is large enough[\cite{JRK1988}]. 

\par
It is worthwhile to mention that the amount of cavity field squeezing in the Jaynes-Cummings model can be enhanced even for low photon number by selective atomic measurements [\cite{CCG1997}]. In addition, the time evolution of the squeezing in the Rabi model [\cite{R1936},\cite{R1937}] was realized numerically for a number of initial states [\cite{PLS1990}]. Besides, by adopting the initial state as a bipartite entangled state consisting of the coherent state in the oscillator subsystem, the squeezing was observed during its evolution [\cite{RC2015},\cite{VY2016}] in the Rabi model. The Rabi model reduces to the familiar Jaynes-Cummings model via the rotating-wave approximation which is solely applicable to the near resonance and weak coupling regime.

\par
However, over recent decades the progress has been made towards the strong coupling regime of the  radiation-matter interactions [\cite{AD2004,GAH2009,TFJ2010,PFE2010,PCB2012,GMD2012,ZFN2013}].  For example, by using circuit quantum electrodynamics the strong coupling of a single photon to a superconducting qubit has been studied experimentally [\cite{AD2004}], the realization of transmission spectra in a superconducting circuit QED system in ultra strong-coupling regime [\cite{TFJ2010}] etc. In addition, experimental observation of the Bloch-Siegert shift [\cite{PFE2010}] also assures the necessity of the counter rotating terms (CRT's) in the description of the Jaynes-Cumming model. This reveals the importance of the CRT's to comprehend the behaviour of full quantum Rabi model for all regimes of the coupling strengths [\cite{CCI2005,JBB2009,JCS2010,Irish2014,PJS2015,RCG2012,SLO2011,PBF2010,Braak2011,You2011,Nataf2011,BDR2012}]. In a recent work [\cite{CJS2017}], qubit-flip-induced cavity mode squeezing in the strong coupling regime of the quantum Rabi model has been investigated. Thus it naturally grows an interest to study the enhancement of squeezing in the Rabi model in presence of a parametric nonlinearity in the strong coupling domain.

\par

To study the qubit-oscillator system under strong interaction where the Hamiltonian includes CRT's, the authors of [\cite{IGMS2005},\cite{AN2010}] introduced an adiabatic approximation scheme that holds in the parameter domain where the oscillator frequency is much larger than the characteristics frequency of the qubit. Based on the separation of different time scales involved in the system, one can reduce the entire dynamics either to qubit or oscillator sector and evaluate the eigenstates of the system approximately [\cite{AN2010}]. To  extend the parameter realm so that it includes both resonance as well as off-resonance, a generalization of the rotating wave approximation has been proposed [\cite{I2007}]. This generalization exploits the basis states obtained in the adiabatic limit and the argument of excitation number conservation according to the rotating wave approximation is also applicable to the Hamiltonian in the new basis. The energy eigenvalues of the resultant block diagonalized Hamiltonian are now approximately valid for strong coupling strengths as well as a wide range of detunings [\cite{I2007}].

\par

Our objective in the present work is as follows. Within the framework of generalized rotating wave approximation, we study the squeezing phenomena in the Rabi model in presence of a  parametric nonlinear term in the strong coupling regime. After approximate diagonalization of the system Hamiltonian, the time evolution of the initial state of the composite system is observed. By tracing over the qubit degree of freedom, we obtain the reduced density matrix corresponding to the oscillator subsystem. This reduced density matrix in turn yields the phase space quasi probability distribution [\cite{WPS2001}] such as Husimi $Q$-function. By exploiting the $Q$-function, we compute the quadrature variance by which the squeezing effect arising in this model is analysed. The work is organised as follows: In Sec. \ref{sec_H}, the approximate diagonalization of the Hamiltonian is performed. In Sec. \ref{dmatrix}, the time evolution of the reduced density matrix corresponding to the oscillator degree of freedom and the $Q$-function is obtained. In Sec. \ref{squeezing}, the squeezing is studied by computing the quadrature variance. Sec. \ref{sec_con} contains the summary of the work.

\section{The approximate diagonalization of the Hamiltonian}
\label{sec_H}
The Rabi Hamiltonian [\cite{FG1937,W1984,P1963,SK1993}] in the presence of a parametric nonlinear term [\cite{S1970,S1974,WE1985,C1997}] can be written as ($\hbar=1$ herein) 
\beq
H = \omega a^{\dag} a + \frac{\Delta}{2} \sigma_{z} 
+ \lambda \sigma_{x} (a^{\dag} + a) + g ({a^{\dag}}^{2} + a^{2}).
\label{RPH}
\eeq
Here, the $(\sigma_{x}, \sigma_{z})$ are Pauli matrices for the qubit having a transition frequency $\Delta$ and the single bosonic mode of frequency $\omega$ is described by the annihilation and creation operators ($a, a^{\dagger}| \hat{n} \equiv a^{\dagger} a$). The coupling between the two subsystems is furnished through a term proportional to $\lambda$ and the constant $g$ corresponds to the strength of the parametric nonlinearity. The Fock states 
$\{\hat{n} |n\rangle = n |n\rangle,\,n = 0, 1,\ldots;\;a \,|n\rangle = \sqrt{n}\,|n - 1\rangle, 
a^{\dagger}\, |n\rangle = \sqrt{n + 1}\,|n + 1\rangle\}$ provide the basis for the oscillator, whereas the eigenstates $\sigma_x |\pm x\rangle = \pm \,|\pm x\rangle$ span the space of the qubit. To obtain the energy spectrum and eigenstates of the Rabi Hamiltonian, numerous approximation schemes have been advanced which are applicable to various ranges of parameters. For instance, to study the dynamical behavior of the qubit-oscillator system we usually employ the well-known rotating wave approximation (RWA) [\cite{JC1963}] since it accurately describes the system in the regime where the oscillator and the qubit frequencies are nearly equal, and also for a weak qubit-oscillator coupling. 

\par
To explore the regimes outside the RWA, an adiabatic approximation scheme [\cite{IGMS2005},\cite{AN2010}] is introduced in the large detuning limit ($\Delta \ll \omega$). To overcome the limitations imposed by the adiabatic approximation which operates only in the large detuning regime, a new method has been proposed [\cite{I2007}] known as the generalized rotating wave approximation that maintains a wide range of validity $(\lambda \sim O(\omega), \Delta \lesssim \omega )$. We adopt the generalized rotating wave approximation to explicitly obtain the approximate eigenenergies and eigenstates for the Hamiltonian (\ref{RPH}) which will be conveniently employed in the study of nonclassical properties emerging from the dynamical evolution of the qubit-oscillator state in a bipartite system. To carry out the generalization of the standard rotating wave approximation, we begin with establishing a new set of basis in the adiabatic approximation which is exploited for representing the Hamiltonian (\ref{RPH}) in the form of a direct sum of $2 \times 2$ blocks along with an entry for uncoupled ground state.
\par
In the adiabatic approximation, diagonalization of the Hamiltonian (\ref{RPH}) is done by considering the qubit's energy splitting $\Delta$ smaller compared to the oscillator's frequency $\omega$, i.e., by allowing the initial energy eigenstate of the oscillator to adiabatically adjusts itself to any changes in the qubit's state $|\pm x\rangle$. Therefore, the qubit's self-energy term can be neglected by choosing $\Delta = 0$ in the Hamiltonian which allows to obtain the oscillator basis. Then, the Hamiltonian (\ref{RPH}) is rewritten and truncated into a $2 \times 2$ block-diagonal form in the aforementioned oscillator basis tensored with the qubit basis. This $2 \times 2$ matrix block structure allows us to compute  eigenenergies and eigenstates of the Hamiltonian which are known as adiabatic- energies and basis of our bipartite system.  Hence, to start with the adiabatic approximation, the oscillator effective Hamiltonian deduced from the Hamiltonian (\ref{RPH}) reads  
\beq
H_{\mathcal{O}}=\omega a^{\dag}a \pm \lambda (a^{\dag} + a) + g ({a^{\dag}}^{2} + a^{2}).
\label{EHO}
\eeq
If $g = 0$, the Hamiltonian $H_{\mathcal{O}}$ is  diagonalizable in the basis $\ket{n_{\pm}}$ containing the degenerate eigenenergies $E_{n} = 
\omega\big(n - \frac{\lambda^{2}}{\omega^{2}}\big)$, where 
the  displaced number states read: $\ket{n_{\pm}}= \mathrm{D}^{\dagger}\left(\pm \frac{\lambda}{\omega}\right) \ket{n},
\, \mathrm{D}\left(\alpha \right) = \exp\left(\alpha a^{\dagger}- \alpha^{*}a \right),\, \alpha \in \mathbb{C}$. Within the adiabatic approximation, the composite state of the system consisting of displaced oscillator basis $\ket{n_{\pm}}$ tensored with the qubit basis $\ket{\pm x}$ are utilized to block-diagonalize the Rabi Hamiltonian which produces non-degenerate eigen spectrum. The overlap between the displaced number states [\cite{IGMS2005}] are given by
\beq
 \braket{m_{-}|n_{+}} =
\begin{cases}
	(-1)^{m-n} \; \left(\frac{2\lambda}{\omega}\right)^{m-n} \; \exp\big(-\frac{2\lambda^{2}}{\omega^{2}}\big) \;
	\sqrt{n!/m!} 
	\; L_n^{(m-n)}(\frac{4\lambda^{2}}{\omega^{2}}), & m \geq n \\
	\left(\frac{2\lambda}{\omega}\right)^{n-m} \; \exp\big(-\frac{2\lambda^{2}}{\omega^{2}}\big) \;
	\sqrt{m!/n!} 
	\; L_m^{(n-m)}(\frac{4\lambda^{2}}{\omega^{2}}) & m < n,
	\label{mn}
\end{cases}
\eeq
where the associated Laguerre polynomial reads $L_{n}^{(j)}(x) = \sum_{k = 0}^{n} 
\,(-1)^{k}\, \binom{n + j}{n - k}\,\frac{x^{k}}{k!}$. The matrix element (\ref{mn}) leads to the identity: 
$\braket{m_{-}|n_{+}} =  (-1)^{n+m} \braket{n_{-}|m_{+}}$. In a similar way, for the case when nonlinear parametric term ($g \neq 0$) is present, we diagonalize the Hamiltonian $H_{\mathcal{O}}$ with the aid of Bogoliubov transformation [\cite{D2019}]. This corresponds to rewriting the Hamiltonian $H_{\mathcal{O}}$ in terms of the new bosonic operators $(\widetilde{a},\widetilde{a}^{\dagger})$
\beq
H_{\mathcal{O}}= \Omega \, \widetilde{a}^{\dagger}\widetilde{a} - \frac{1}{2}(\omega-\Omega)
- \frac{\lambda^{2}}{\omega+2g},
\label{BOH}
\eeq
where the operators $(\widetilde{a},\widetilde{a}^{\dagger})$ obeying the standard bosonic commutation relations are represented as
\beq
\widetilde{a} = \mathrm{S}^{\dag}(r) \mathrm{D}^{\dag}(\eta) a \mathrm{D}(\eta) \mathrm{S}(r), \;
\widetilde{a}^{\dag} = \mathrm{S}^{\dag}(r) \mathrm{D}^{\dag}(\eta) a^{\dag} \mathrm{D}(\eta) \mathrm{S}(r).
\label{BO}
\eeq
The squeezing operator given in (\ref{BO}) reads, $\mathrm{S}(\xi)=\exp((\xi {a^{\dag}}^{2}- \xi^{*} a^{2})/2)$, $\xi= r \exp(i \vartheta)$, $\xi \in \mathbb{C}$, and it maintains the following unitary transformations:
\beq
\mathrm{S}^{\dag}(\xi) a \mathrm{S}(\xi) = \mu a + \nu a^{\dag}, \quad \mathrm{S}^{\dag}(\xi) a^{\dag} \mathrm{S}(\xi) = \mu a^{\dag} + \nu^{*} a, \quad \mu = \cosh(r), \; \nu = \exp(i \vartheta) \sinh(r),
\label{sqo}
\eeq
where we denote the abbreviations: $\Omega = \sqrt{\omega^{2}-4g^{2}}$, $r=arc \cosh \left( \sqrt{\frac{\omega + \Omega}{2 \Omega}} \right)$ and $\eta = \sqrt{\frac{\omega + \Omega}{2 \Omega}}  \left(1+ \frac{\omega - \Omega}{2g} \right) \frac{\lambda}{\omega + 2g} $. Now, the effective Hamiltonian $H_{\mathcal{O}}$ (\ref{BOH}) can be diagonalized in the following oscillator basis $\ket{r,n_{\pm}}$
\beq
H_{\mathcal{O}} \ket{r,n_{\pm}} = E_{n} \ket{r,n_{\pm}}, \quad
E_{n}=(n+\frac{1}{2})\Omega - \frac{\omega}{2} - \frac{\lambda^{2}}{\omega + 2g}, \quad
\ket{r,n_{\pm}} = \mathrm{S}^{\dag}(r)\mathrm{D}^{\dag}(\pm \eta) \ket{n}.
\label{AOE}
\eeq
Thereafter, by utilizing the oscillator basis (\ref{AOE}) tensored with the qubit basis: $\ket{\pm x ; r ,n_{\pm}}\equiv \ket{\pm x} \otimes \ket{ r ,n_{\pm}}$, the Hamiltonian (\ref{RPH}) is truncated into $2 \times 2$ blocks
\beq
\begin{pmatrix}
	E_{n} & \Delta_{n}   \\
	\Delta_{n}  & E_{n}
\end{pmatrix}, \; \Delta_{n} = \frac{\Delta}{2} \exp(-2 \eta^{2}) L_{n}(4 \eta^{2}), \; n\geq 0.
\label{2X2AD}
\eeq
From the above matrix representation (\ref{2X2AD}), the adiabatic- energies and the basis are obtained:
\beq
E_{\pm,n }=E_{n} \pm \Delta_{n}, \quad \ket{E_{\pm,n}} = \frac{1}{\sqrt{2}}( \ket{x;r,n_{+}} \pm \ket{-x;r,n_{-}} ).
\label{ADE}
\eeq
\begin{figure}
	\begin{center}
		\captionsetup[subfigure]{labelformat=empty}
		\subfloat[$(\mathsf{a}_{1})$]{\includegraphics[width=5.5cm,height=6cm]{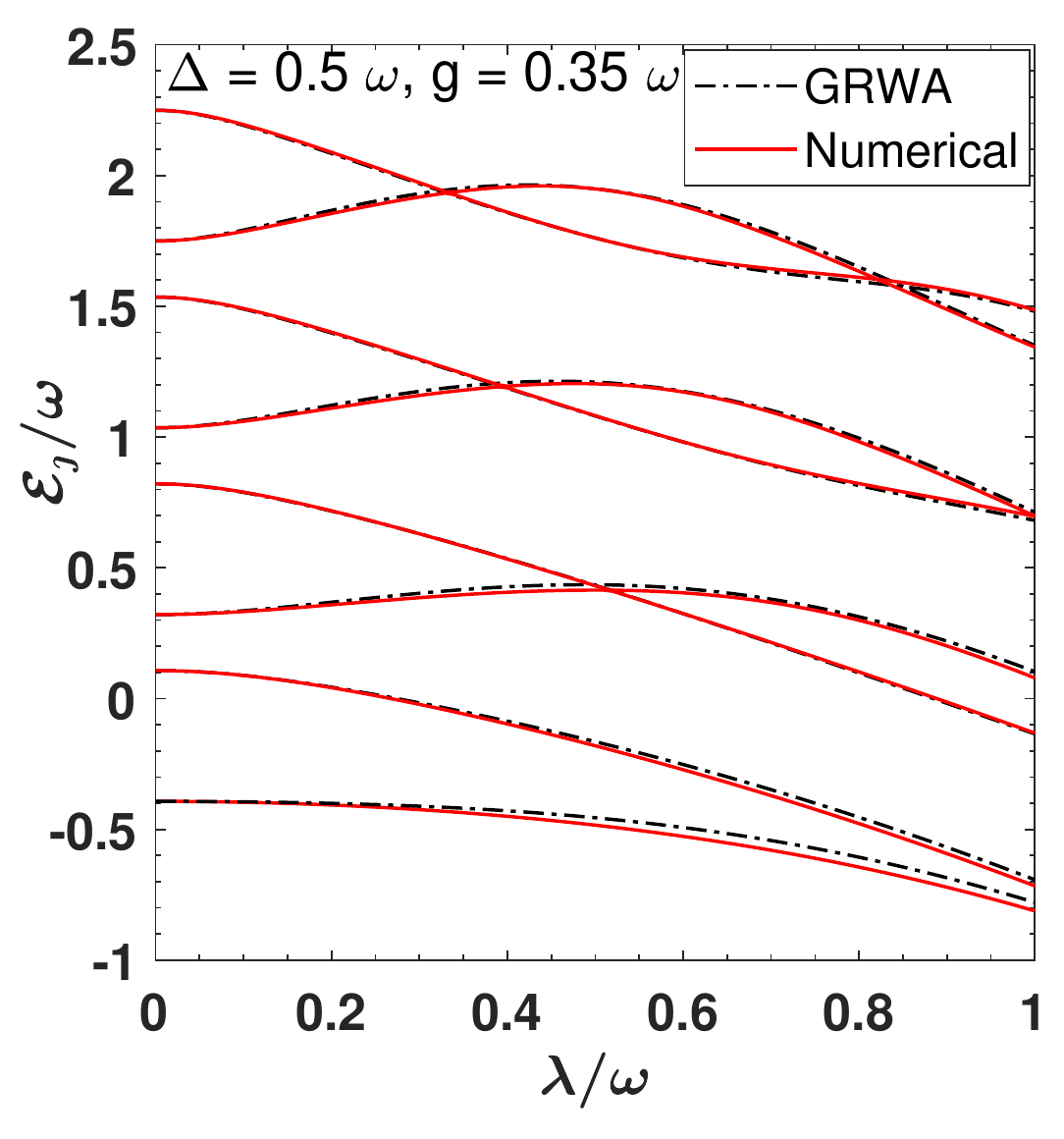}} 
		\captionsetup[subfigure]{labelformat=empty}
		\subfloat[$(\mathsf{a}_{2})$]{\includegraphics[width=5.5cm,height=6cm]{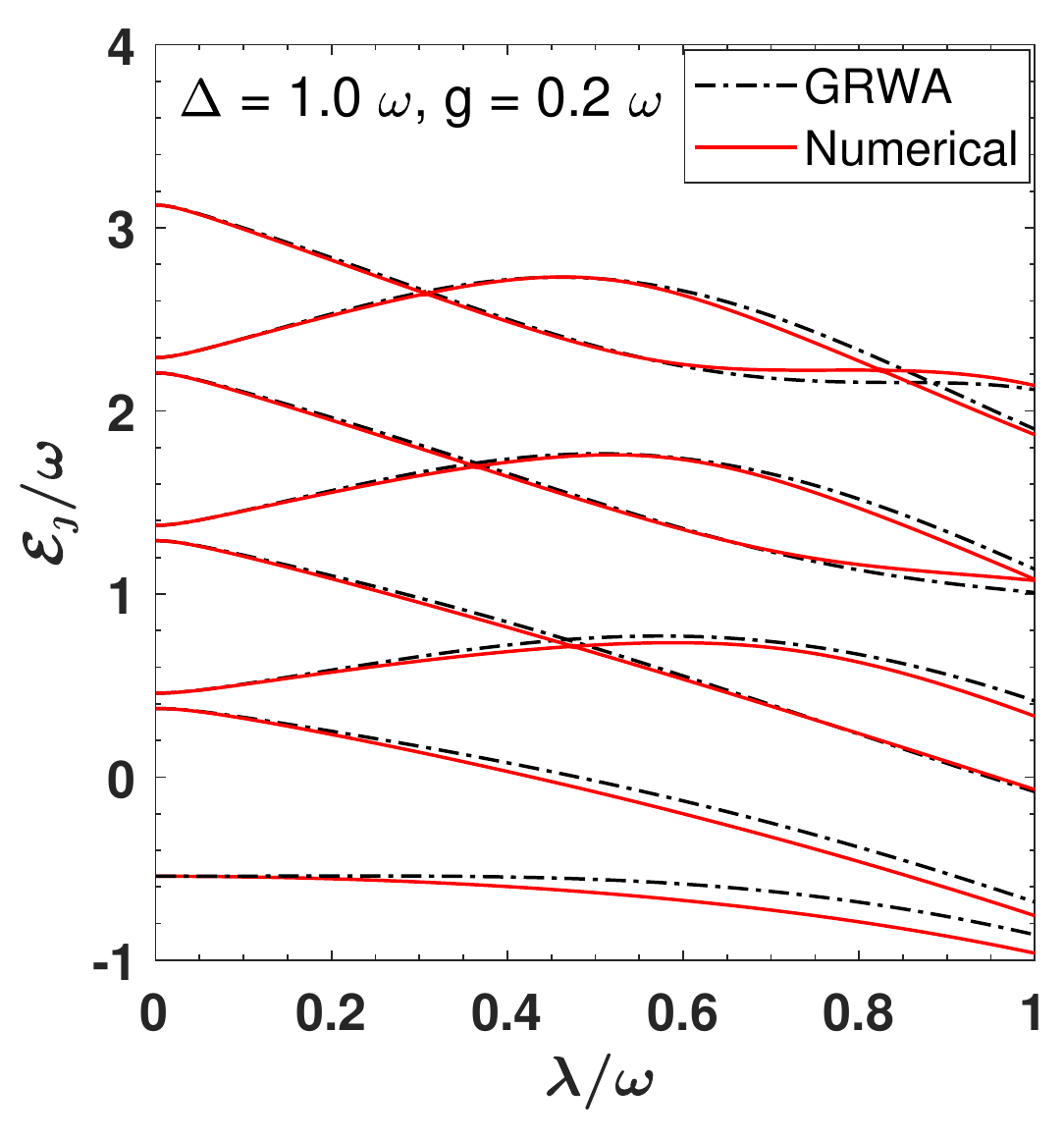}} 
		\caption{(Color online). Generalized rotating wave approximation (GRWA) energy levels (\ref{energy_GRWA}) (dotted-dashed) are compared with numerically-determined (solid) energies as a function of coupling strength $\lambda/\omega$  $\mathsf{(a_1)}$ for the parameter values  $g = 0.35 \;\omega $  in the off-resonance $(\Delta = 0.5\, \omega)$ and $\mathsf{(a_2)}$ in the resonance $(\Delta =1.0 \, \omega)$ with  $g=0.2 \,\omega$. Our GRWA approach works well at far away from the resonance say $\Delta \lesssim 0.5 \,$ in the parameters regime $\lambda \lesssim 1.0 \, \omega$ and $g \lesssim 0.35 \, \omega$. Similarly in the resonance, the admissible regime of the parameter $g \lesssim 0.2 \, \omega$ which is evident from $\mathsf{(a_2)}$. }  
		\label{Fig_Eng_lam}
	\end{center}
\end{figure}
\indent Furthermore, the adiabatic basis (\ref{ADE}) is exploited towards approximate diagonalization of the Hamiltonian (\ref{RPH}) via the generalized rotating wave approximation [\cite{I2007}] and the resulting matrix elements truncated into $2 \times 2$ blocks apart from the uncoupled ground state which can be written as
\beq
\begin{pmatrix}
	E_{+,n-1} & \tilde{\Delta}_{n}   \\
	\tilde{\Delta}_{n}  & E_{-,n}
\end{pmatrix}, \; \tilde{\Delta}_{n} = \frac{ \eta \Delta}{ \sqrt{n}} \exp \left(- 2\eta^{2}\right) L^{(1)}_{n-1} \left( 4 \eta^{2} \right), \; n \geq 1.
\label{2X2GR}
\eeq
The  uncoupled ground state energy and solutions for the simple block-diagonal form (\ref{2X2GR}) of the doublets explicitly read
\bea
\mathcal{E}_{0} & \equiv & E_{-,0} =\frac{\Omega - \omega}{2}-\frac{\lambda^{2}}{\omega + 2g}- \frac{\Delta}{2} \exp(-2 \eta^{2}),\nn\\ 
\mathcal{E}_{\pm,n (\ge 1)}& = & \Big(n \Omega - \frac{\omega}{2} - \frac{\lambda^{2}}{\omega+2g}\Big) 
+ \frac{\Delta}{4} \exp(-2 \eta^{2}) 
\Big(L_{n-1} (4 \eta^{2}) - L_{n}(4 \eta^{2})\Big) \nn \\
& \pm &\frac{1}{2}\sqrt{ \Big \lgroup  \Omega - \frac{\Delta}{2} \exp(-2 \eta^{2})
	\Big( L_{n-1} (4 \eta^{2}) + L_{n} (4 \eta^{2}) \Big) \Big \rgroup^2
	+ \frac{ \eta^{2} \Delta^2 }{n} \exp(-4\eta^{2}) \Big \lgroup L_{n-1}^{(1)}(4 \eta^{2}) \Big \rgroup^2}. \qquad
\label{energy_GRWA}
\eea
The corresponding eigenstates are given by
\bea
|{\mathcal E}_0 \rangle &\equiv& \ket{{E}_{-,0}} = \dfrac{1}{\sqrt{2}} 
\Big( \ket{x;r,0_{+}} - \ket{-x;r,0_{-}} \Big),\nn \\ 
|{\mathcal E}_{\pm,n (\ge 1)} \rangle &=& \zeta_{\pm,n}  | E_{+,{n-1}} 
\rangle \pm \dfrac{\tilde{\Delta}_{n}}{|\tilde{\Delta}_{n}|} \zeta_{\mp,n} |E_{-,n} \rangle,
\quad   \zeta_{\pm,n} = \sqrt{\frac{\chi_n \pm \varepsilon_{n}} {2\chi_n}},
\label{Hn_eigenstate}
\eea
here we  abbreviate: $\chi_n=\sqrt{{\tilde{\Delta}_{n}}^2+\varepsilon_{n}^2},\,\varepsilon_{n} 
=\frac{E_{+,n-1}-E_{-,n}}{2}$. The completeness relation of the orthonormal bipartite basis 
(\ref{Hn_eigenstate}) now reads: 
\beq
|{\mathcal E}_{0} \rangle \langle {\mathcal E}_{0}| + \sum_{n = 1}^{\infty} 
\left(|{\mathcal E}_{+,n} \rangle \langle {\mathcal E}_{+,n}| +  |{\mathcal E}_{-,n} \rangle 
\langle {\mathcal E}_{-,n}|\right) = \sum_{n = 0}^{\infty} 
\left(| E_{+,n} \rangle \langle  E_{+,n}| +  |E_{-,n} \rangle 
\langle E_{-,n}|\right) = \mathbb{I}.
\label{completeness}
\eeq  
\section{Time evolution of the oscillator's reduced density matrix and the Husimi Q-function}
\setcounter{equation}{0}
\label{dmatrix}
Upon completion of the above construction of the energy eigenstates $(\ref{Hn_eigenstate})$ via the generalized rotating wave approximation, we further proceed to explore the role of parameter $g$ on the nonclassical features of the oscillator degree of freedom, in particular, the Husimi $Q$-function and the squeezing through the dynamics of the bipartite system. The initial state of the qubit-oscillator system reads: $\ket{\psi(0)} = \ket{-x} \otimes \ket{0}$, where $\ket{0}$ is the vacuum state of the oscillator. The time evolution of the initial state is
\beq 
\ket{\psi(t)} = \mathcal{C}_{0} (t) \ket{\mathcal{E}_{0}} + \sum_{n=1}^{\infty} \mathcal{C}_{\pm,n} (t) \ket{\mathcal{E}_{\pm,n}}, \quad \mathcal{C}_{0}(t) = \mathcal{C}_{0} \exp(-i \mathcal{E}_{0} t),\quad \mathcal{C}_{\jmath, n}(t) = \mathcal{C}_{\jmath, n} \exp(-i \mathcal{E}_{\jmath} t), \quad \jmath \in \{\pm \},
\label{wave_time}
\eeq
where the coefficients read:
\bea
\mathcal{C}_{0} \!\! &=& \!\! -\frac{1}{\sqrt{2 \mu}} \exp \left( -\frac{\eta^{2}}{2} + \frac{\nu \eta^{2}}{2 \mu} \right), \; \frac{\nu}{\mu} = \frac{\omega - \Omega}{2g}, \nn \\
\mathcal{C}_{\pm,n} \!\! &=& \!\! -\mathcal{C}_{0} \left( \frac{-\nu}{2 \mu}\right)^{\!\! \frac{n}{2}} 
\left( \frac{\zeta_{\pm,n}}{\sqrt{(n-1)!}} \left( \frac{-\nu}{2 \mu}\right)^{\!\! -\frac{1}{2}} 
\mathrm{H}_{n-1}\left( \frac{i(\mu-\nu)\eta}{\sqrt{2 \mu \nu}} \right) \mp 
\dfrac{\tilde{\Delta}_{n}}{|\tilde{\Delta}_{n}|} \frac{\zeta_{\mp,n}}{\sqrt{n!}} \;
\mathrm{H}_{n}\left( \frac{i(\mu-\nu)\eta}{\sqrt{2 \mu \nu}} \right) \right),
\label{coeff}
\eea
here the Hermite polynomials are given by the exponential generating function [\cite{gradshteyn2007}]: $\exp(2 \,\mathsf{x} \mathsf{t}-\mathsf{t}^{2}) = \sum_{n=0}^{\infty} \frac{\mathrm{H}(\mathsf{x}) \mathsf{t}^{n}}{n!}$. To facilitate the construction of the time evolution of the initial state (\ref{wave_time}), we provide the following expansion of squeezed coherent state in the number state basis [\cite{SZ2001}] together with the property below:
\beq
\mathrm{S}(\xi) \mathrm{D}(\alpha) \ket{0} = \exp \left( -\frac{|\alpha|^{2}}{2} - \frac{\alpha^{2} \nu^{*}}{2 \mu}\right)
\sum_{n=0}^{\infty} \frac{i^{n}}{\sqrt{n! \mu}} \left( \frac{\nu}{2 \mu}\right)^{\!\! \frac{n}{2}}
\mathrm{H}_{n}\left( \frac{-i\alpha}{\sqrt{2 \mu \nu}} \right) \ket{n}, \, 
 \mathrm{D}(\alpha) \mathrm{S}(\xi) = \mathrm{S}(\xi) \mathrm{D}(\alpha \mu - \alpha^{*} \nu).
\label{def_sq}
\eeq
The above expressions are utilized to compute the coefficients of $\ket{\psi(t)}$ given in (\ref{coeff}). The normalization of the state $\ket{\psi(t)}$: $\braket{\psi(t)|\psi(t)} \equiv |\mathcal{C}_{0}(t)|^{2}+\sum_{n=1}^{\infty} |\mathcal{C}_{\pm,n}(t)|^{2}=1$ can be shown by exploiting the following identity [\cite{AAR1999}]
\beq
\sum_{n=0}^{\infty}\dfrac{\mathsf{t}^{n}}{2^{n}n!} \mathrm{H}_{n}(\mathsf{x}) \mathrm{H}_{n}(\mathsf{y}) =
\dfrac{1}{\sqrt{1-\mathsf{t}^{2}}}\,
\exp\left(-\dfrac{(\mathsf{tx})^{2}-2\mathsf{txy}+
	(\mathsf{ty})^{2}}{ 1-\mathsf{t}^{2}} \right).
\label{Hermite_id} 
\eeq

Therefore, the time evolution of the density matrix of the bipartite pure state can be represented as
\beq
 \rho(t) \equiv \ket{\psi(t)} \bra{\psi(t)}.
\eeq
\par
The reduced density matrix for the oscillator is obtained by partial tracing over the qubit-Hilbert space i.e. 
 \indent $\rho_{\mathcal{O}} \equiv \mathrm{Tr}_{\mathcal{Q}} \rho$. Its explicit construction reads:
\bea
\rho_{\mathcal{O}}(t) \! \! \! &=& \! \! \! |\mathcal{C}_{0}(t)|^{2} P_{0,0}^{(+)} \! + \! 
\sum_{n=1}^{\infty} \Big( \mathcal{C}_{0}(t) {\mathcal{A}_{n}(t)}^{*} P_{0,n-1}^{(-)}  \! + \!
{\mathcal{C}_{0}(t)}^{*} \mathcal{A}_{n}(t) P_{n-1,0}^{(-)} 
+ \mathcal{C}_{0}(t) {\mathcal{B}_{n}(t)}^{*} P_{0,n}^{(+)} \; \; \; \nn \\ 
& &+ {\mathcal{C}_{0}(t)}^{*} \mathcal{B}_{n}(t) P_{n,0}^{(+)} \Big) +
\sum_{n,m=1}^{\infty} \Big( \mathcal{A}_{n}(t) {\mathcal{A}_{m}(t)}^{*} P_{n-1,m-1}^{(+)} 
+ \mathcal{B}_{n}(t) {\mathcal{B}_{m}(t)}^{*} P_{n,m}^{(+)} \nn \\
& &+ \mathcal{B}_{n}(t) {\mathcal{A}_{m}(t)}^{*} P_{n,m-1}^{(-)} +
\mathcal{A}_{n}(t) {\mathcal{B}_{m}(t)}^{*} P_{n-1,m}^{(-)} \Big), \nn \\
\mathcal{A}_{n}(t) &=& \zeta_{+,n} \, \mathcal{C}_{+,n}(t) + \zeta_{-,n} \, \mathcal{C}_{-,n}(t),\quad
\mathcal{B}_{n}(t) = \frac{\tilde{\Delta}_{n}}{|\tilde{\Delta}_{n}|} 
\left( \zeta_{-,n} \mathcal{C}_{+,n}(t) - \zeta_{+,n} \mathcal{C}_{-,n}(t) \right),
\label{density_oscillator}
\eea
where the projection operators read
$P_{n,m}^{(\pm)}= \frac{1}{2} \left( \ket{r,n_{+}}\bra{r,m_{+}} \pm \ket{r,n_{-}}\bra{r,m_{-}} \right),\,(n, m = 0, 1, \ldots)$. The 
density matrix (\ref{density_oscillator}) obeys the normalization condition:
$\mathrm{Tr}\,\rho_{\mathcal{O}}(t) = 1$.
\par
The Husimi $Q$-function [\cite{WPS2001}] is a quasi probability distribution  defined  as expectation value of the oscillator density matrix 
in an arbitrary coherent state. It assumes nonnegative values on the phase space in contrast to the other phase space quasi probability distributions. Being easily 
computable it has been extensively used [\cite{SA2002}, \cite{IWAH2003}] in the study of the 
occupation on the phase space. 
For our reduced density matrix of the oscillator (\ref{density_oscillator}), the corresponding $Q$-function reads
\beq
Q(\beta,\beta^{*}) = \frac{1}{\pi} \bra\beta\rho_{\cal O}\ket\beta, \; \ket{\beta} = \mathrm{D}(\beta) \ket{0}, \; \beta \in \mathbb{C}.
\label{Q_defn}
\eeq 
Our construction of the oscillator density matrix (\ref{density_oscillator}) now yields the time-evolution of the $Q$-function:
\bea
Q(\beta,\beta^{*}) &=& \dfrac{1}{2}  |\mathcal{C}_{0}(t)|^{2} H^{(+)}_{0,0}(\beta,\beta^{*})
+ \mathrm{Re} \Big \lgroup \mathcal{C}_{0}(t)^{*} 
\sum_{n=1}^{\infty} \Big( \mathcal{A}_{n}(t) H^{(-)}_{0,n-1}(\beta,\beta^{*})\nn  \\
&+& \mathcal{B}_{n}(t) H^{(+)}_{0,n}(\beta,\beta^{*}) \Big) + 
\sum_{n,m=1}^{\infty} \! \mathcal{A}_{n}(t)^{*} 
\mathcal{B}_{m}(t) H^{(-)}_{n-1,m}(\beta,\beta^{*}) \Big \rgroup \nn \\
&+&\dfrac{1}{2}\! \sum_{n,m=1}^{\infty} \!\Big( \mathcal{A}_{n}(t)^{*} 
\mathcal{A}_{m}(t) H^{(+)}_{n-1,m-1}(\beta,\beta^{*}) 
+\mathcal{B}_{n}(t)^{*} \mathcal{B}_{m}(t) H^{(+)}_{n,m}(\beta,\beta^{*}) \Big).
\label{Q_function}
\eea
Here, the weight functions on the phase space read
\bea
H^{(\pm)}_{n,m} (\beta, \beta^{^*}) \!\!\!\! &=& \!\!\!\! \dfrac{(-1)^{n}}{\pi \mu \sqrt{n!m!}} 
\left( - \frac{\nu}{2 \mu}\right)^{\!\! \frac{n+m}{2}}
\exp \left(- \frac{\eta^{2}}{\mu} \left( \left( \mu - \nu \right) + \frac{4\mu}{\nu^{2}} \right) \right) 
\exp \left(- \Big| \beta + \frac{\nu}{2 \mu} \beta^{*} \Big|^{2} \right) \times \qquad \qquad \qquad \nn \\
&\times& \!\!\!\! \Bigg \lgroup \exp \left( \Big| \frac{\nu \beta}{2 \mu} - \frac{2\eta }{\nu}  \Big|^{2} \right) \mathrm{H}_{n}\left(i \frac{\beta_{+}^{*}}{\sqrt{2 \mu \nu}}\right) 
\mathrm{H}_{m}\left(-i \frac{\beta_{+}}{\sqrt{2 \mu \nu}}\right) \pm 
\exp \left( \Big| \frac{\nu \beta}{2 \mu} + \frac{2\eta }{\nu}  \Big|^{2} \right) \times \nn \\
&\times& \!\!\!\! \mathrm{H}_{n}\left(i \frac{\beta_{-}^{*}}{\sqrt{2 \mu \nu}}\right) 
\mathrm{H}_{m}\left(-i \frac{\beta_{-}}{\sqrt{2 \mu \nu}}\right)
 \Bigg \rgroup,
\label{Q_weight}
\eea
where $\beta_{\pm}=\beta \pm \eta (\mu - \nu)$. To arrive at the expression (\ref{Q_function}) we make use of the following inner products
\beq
\braket{\beta|\xi,n_{\pm}} = \frac{1}{\sqrt{\mu n!}} \left(-i \sqrt{\frac{\nu^{*}}{2\mu}} \right)^{n}
\exp \left(- \frac{|\alpha|^{2}}{2} + \frac{\alpha^{2}\nu^{*}}{2\mu} 
- \frac{|\beta|^{2}}{2} - \frac{\beta^{*2}\nu}{2\mu} - \frac{\alpha \beta^{*}}{\mu}\right)
\mathrm{H}_{n}\left(  \frac{i (\mu \alpha^{*}- \alpha \nu^{*}+ \beta^{*})}{\sqrt{2 \mu \nu^{*}}}\right),
\eeq
with $\braket{\beta|\xi,n_{\pm}} \equiv \braket{0|D^{\dagger}(\beta) S^{\dagger}(\xi)  D^{\dagger}(\pm \alpha)|n}$, and these inner products can be calculated by utilizing the expressions (\ref{def_sq}). The expression (\ref{Q_function}) can be shown to satisfy the normalization criteria i.e. $\int Q(\beta, \beta^{*}) \mathrm{d}^{2}\beta = 1$ 

by employing the following integrals
\bea
\int \mathrm{d}^{2}\beta \exp\left(-|\beta|^{2}-\frac{\nu}{2\mu}(\beta^{2}+\beta^{*2}) 
\mp \frac{\eta}{\mu} (\beta + \beta^{*}) \right) \!\!\!\!\! \!\!\!\!\! & & \!\!\!\!\! \mathrm{H}_{n}\left(i \frac{\beta_{\pm}^{*}}{\sqrt{2 \mu \nu}}\right) 
\mathrm{H}_{m}\left(-i \frac{\beta_{\pm}}{\sqrt{2 \mu \nu}}\right) \nn \\
&=&  \pi\, \mu \, n! \left( \frac{2 \mu}{\nu}\right)^{n} 
\exp \left( \frac{\eta^{2}}{\mu} (\mu - \nu)\right) \, \delta_{n,m} \, ,
\eea
and it also maintains the bounds: $0 \leq Q(\beta,\beta^{*}) \leq \frac{1}{\pi}$. 
\begin{figure}
	\begin{center}
		\captionsetup[subfigure]{labelformat=empty}
		\subfloat[$(\mathsf{a}_{1})$]{\includegraphics[width=4.5cm,height=4.5cm]{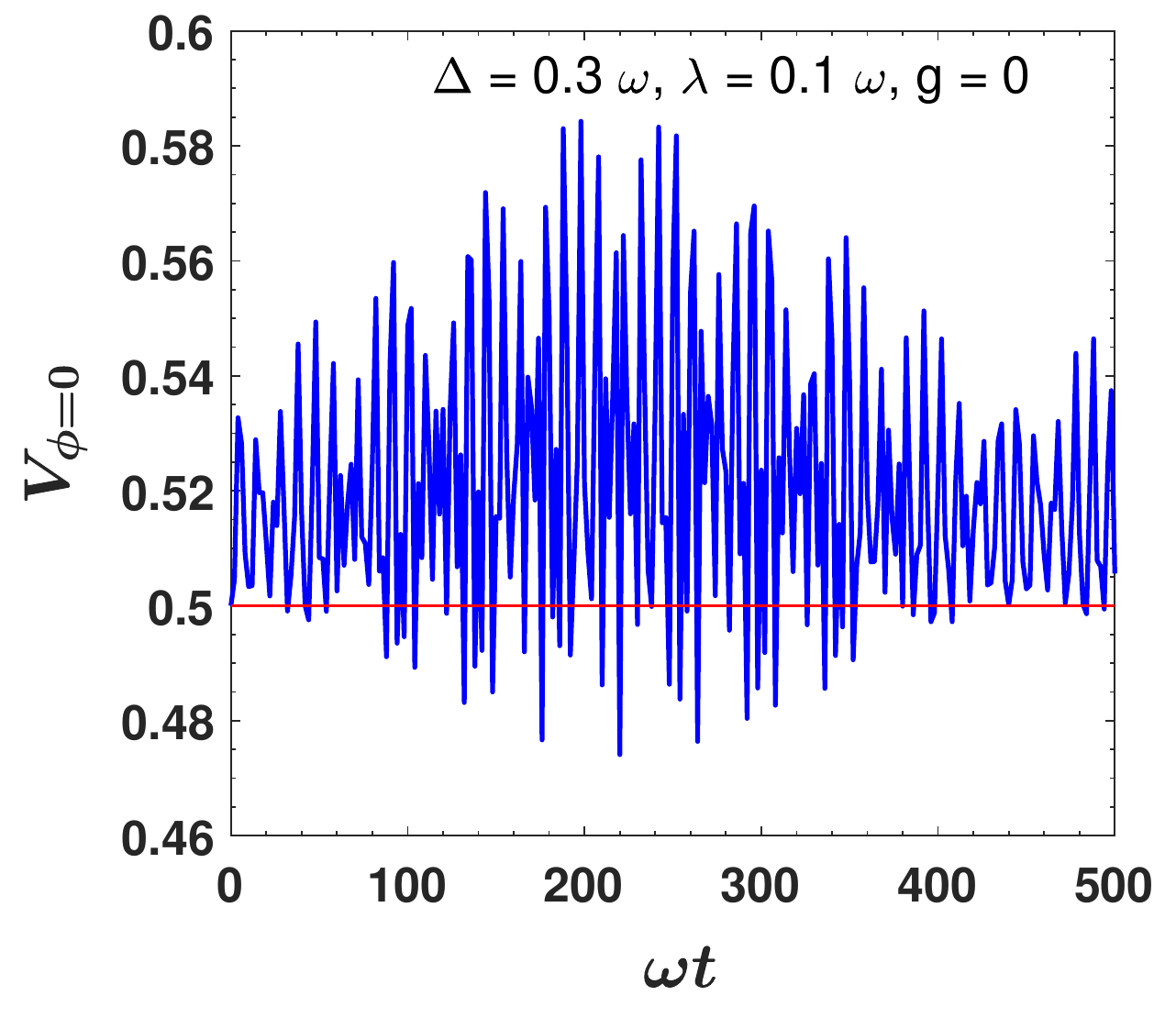}} 
		\captionsetup[subfigure]{labelformat=empty}
		\subfloat[$(\mathsf{a}_{2})$]{\includegraphics[width=3.5cm,height=3.5cm]{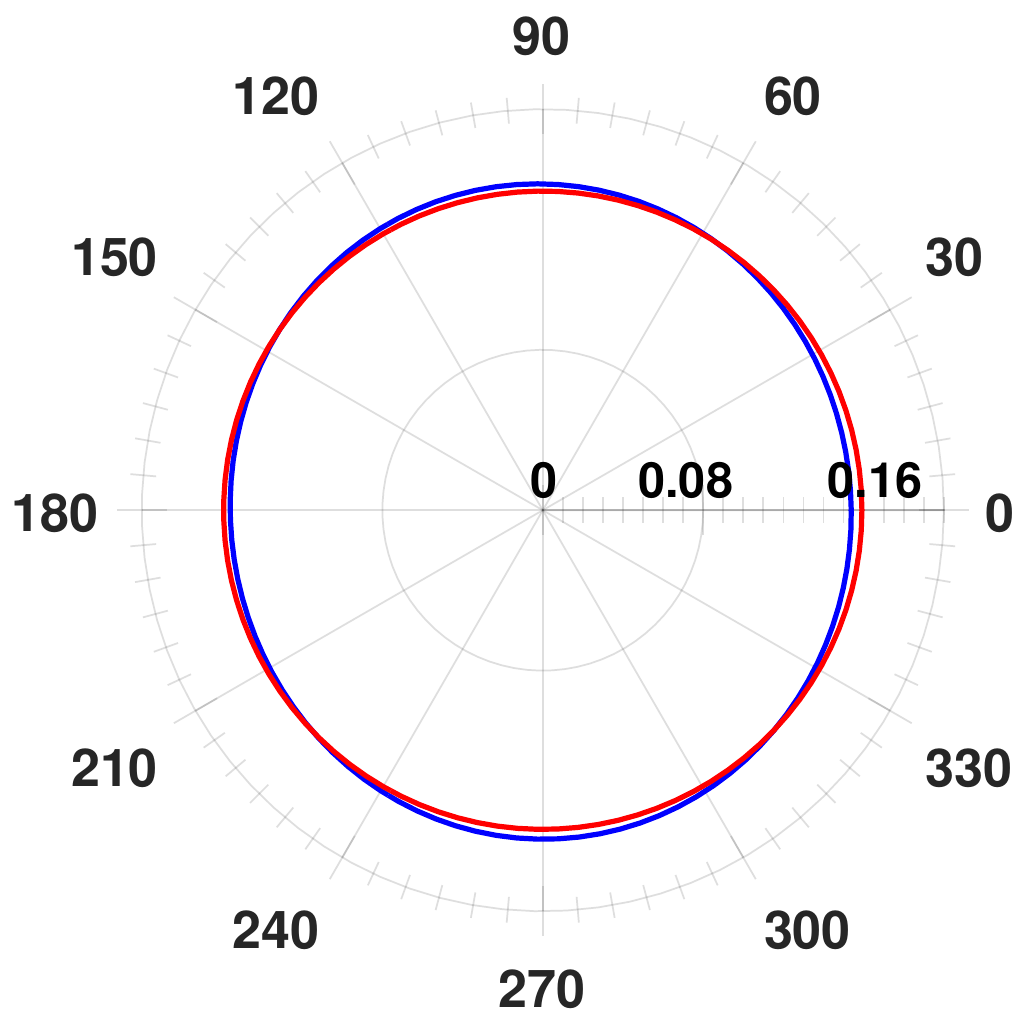}} 
		\captionsetup[subfigure]{labelformat=empty}
		\subfloat[$(\mathsf{a}_{3})$]{\includegraphics[width=3.5cm,height=3.5cm]{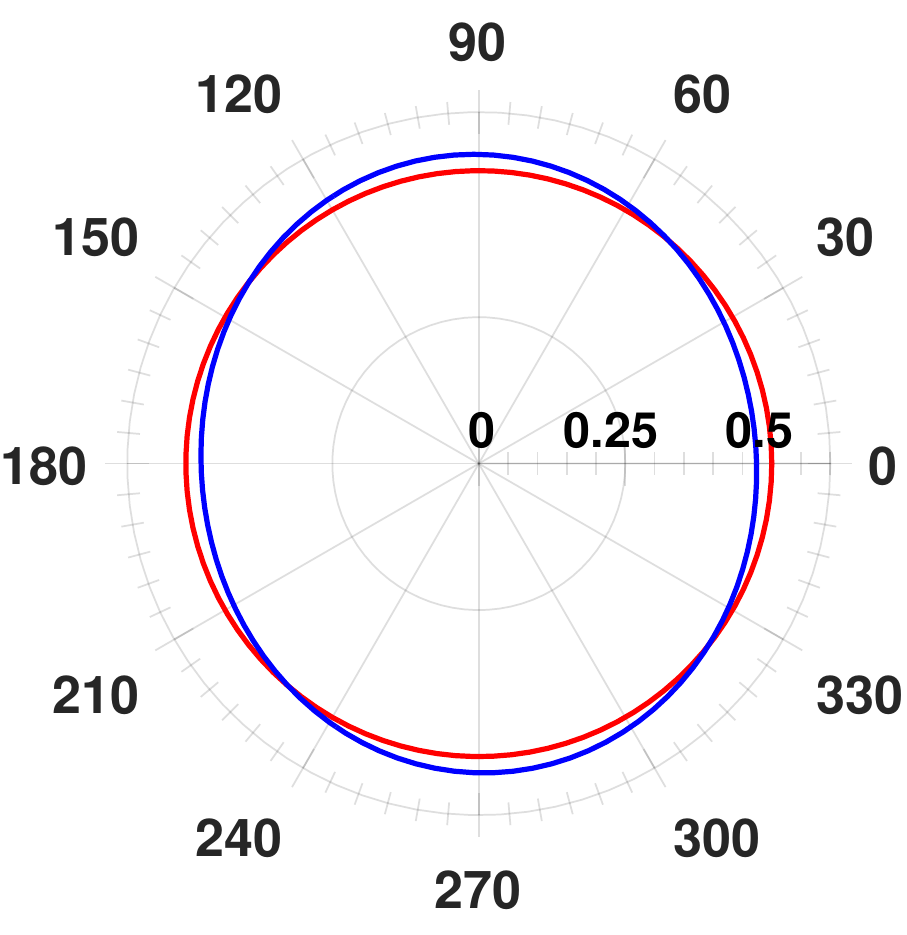}} 
		\captionsetup[subfigure]{labelformat=empty}
		\captionsetup[subfigure]{labelformat=empty}
		\subfloat[$(\mathsf{a}_{4})$]{\includegraphics[width=4.5cm,height=4.5cm]{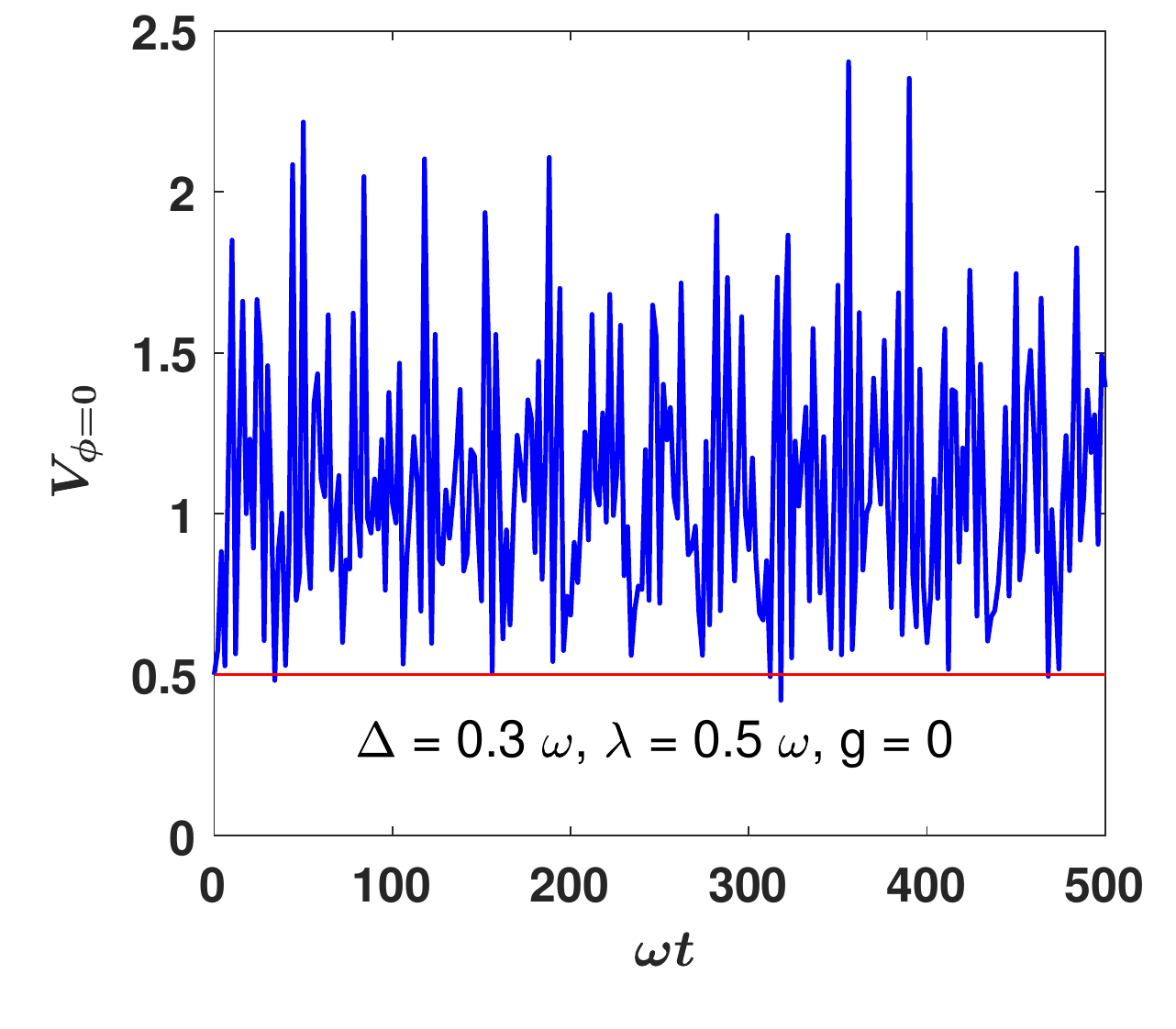}} 
		\captionsetup[subfigure]{labelformat=empty}
		\subfloat[$(\mathsf{b}_{1})$]{\includegraphics[width=4.5cm,height=4.5cm]{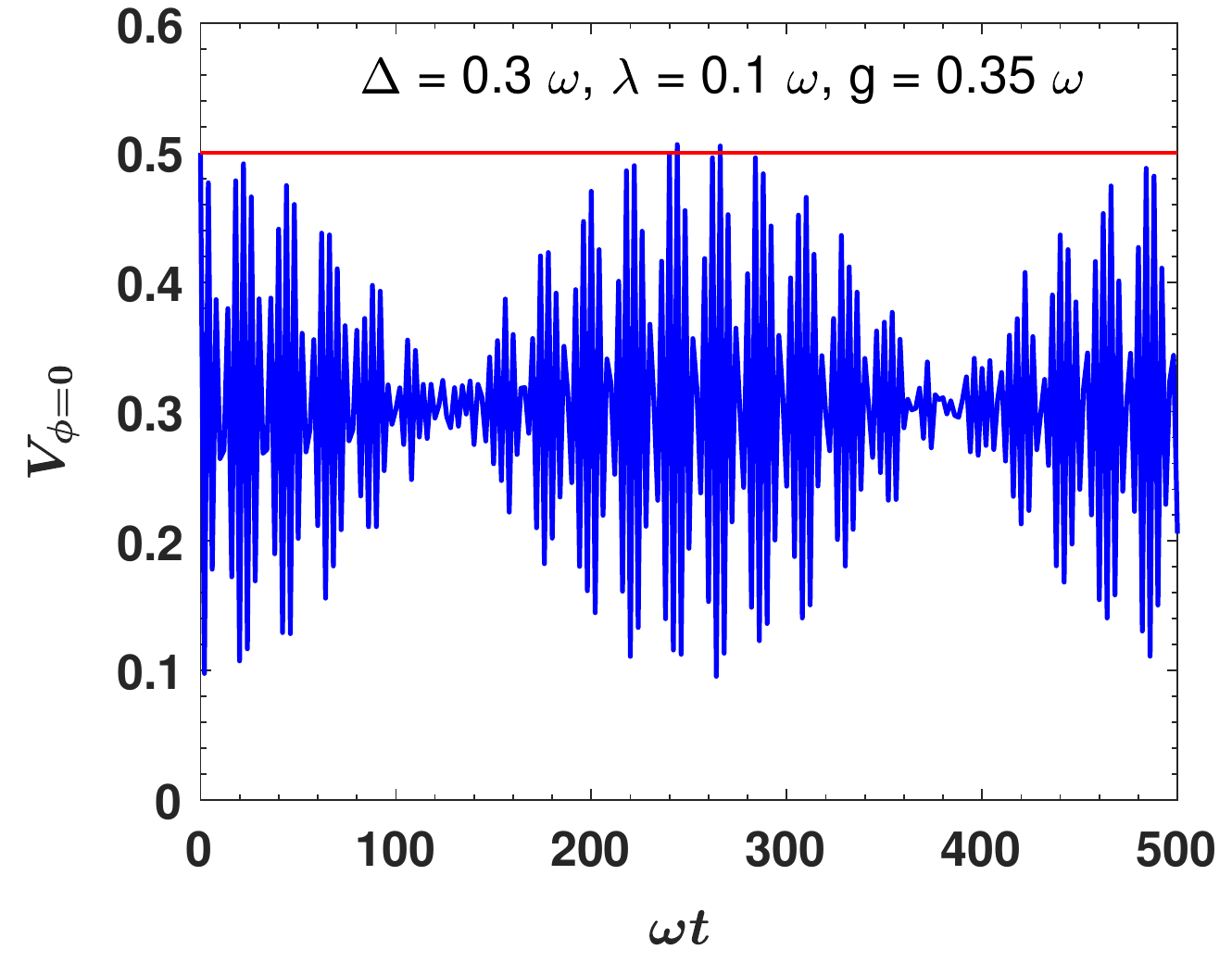}} 
		\captionsetup[subfigure]{labelformat=empty}
		\subfloat[$(\mathsf{b}_{2})$]{\includegraphics[width=3.5cm,height=3.5cm]{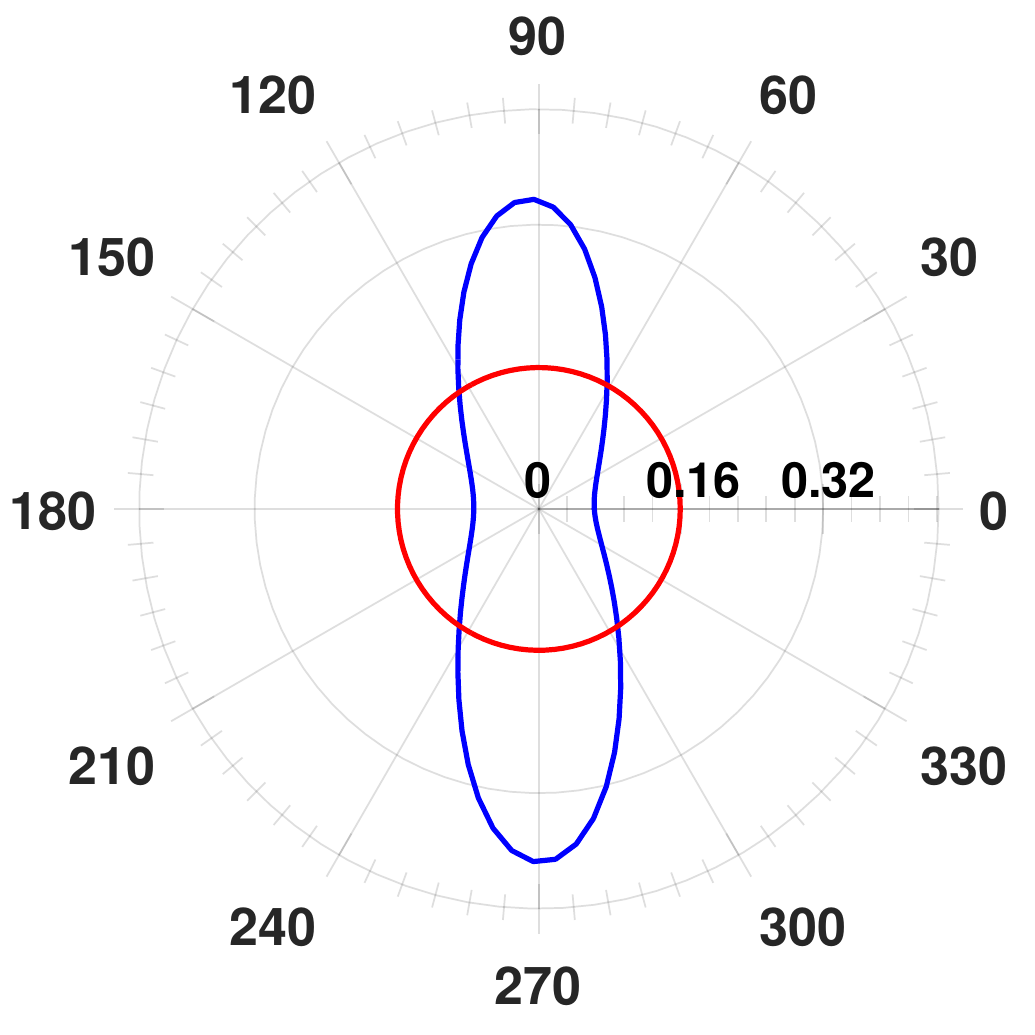}} 
		\captionsetup[subfigure]{labelformat=empty}
		\subfloat[$(\mathsf{b}_{3})$]{\includegraphics[width=3.5cm,height=3.5cm]{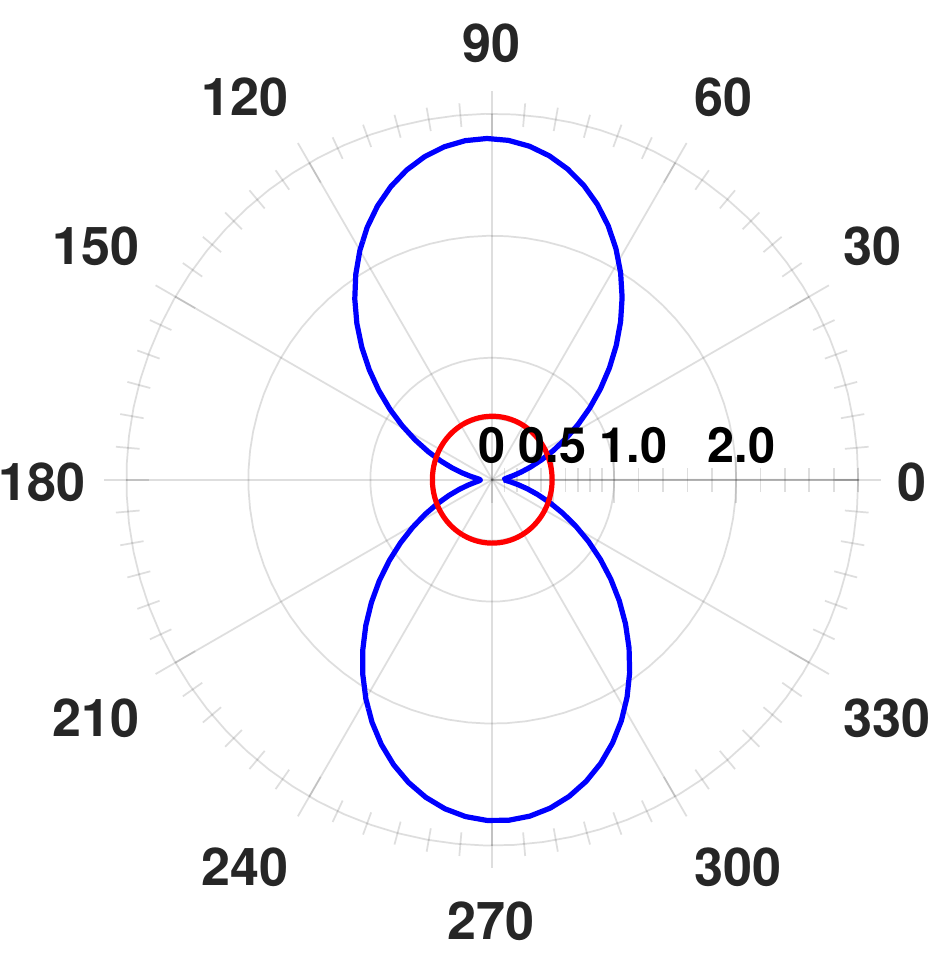}} 
		\captionsetup[subfigure]{labelformat=empty}
		\subfloat[$(\mathsf{b}_{4})$]{\includegraphics[width=4.5cm,height=4.5cm]{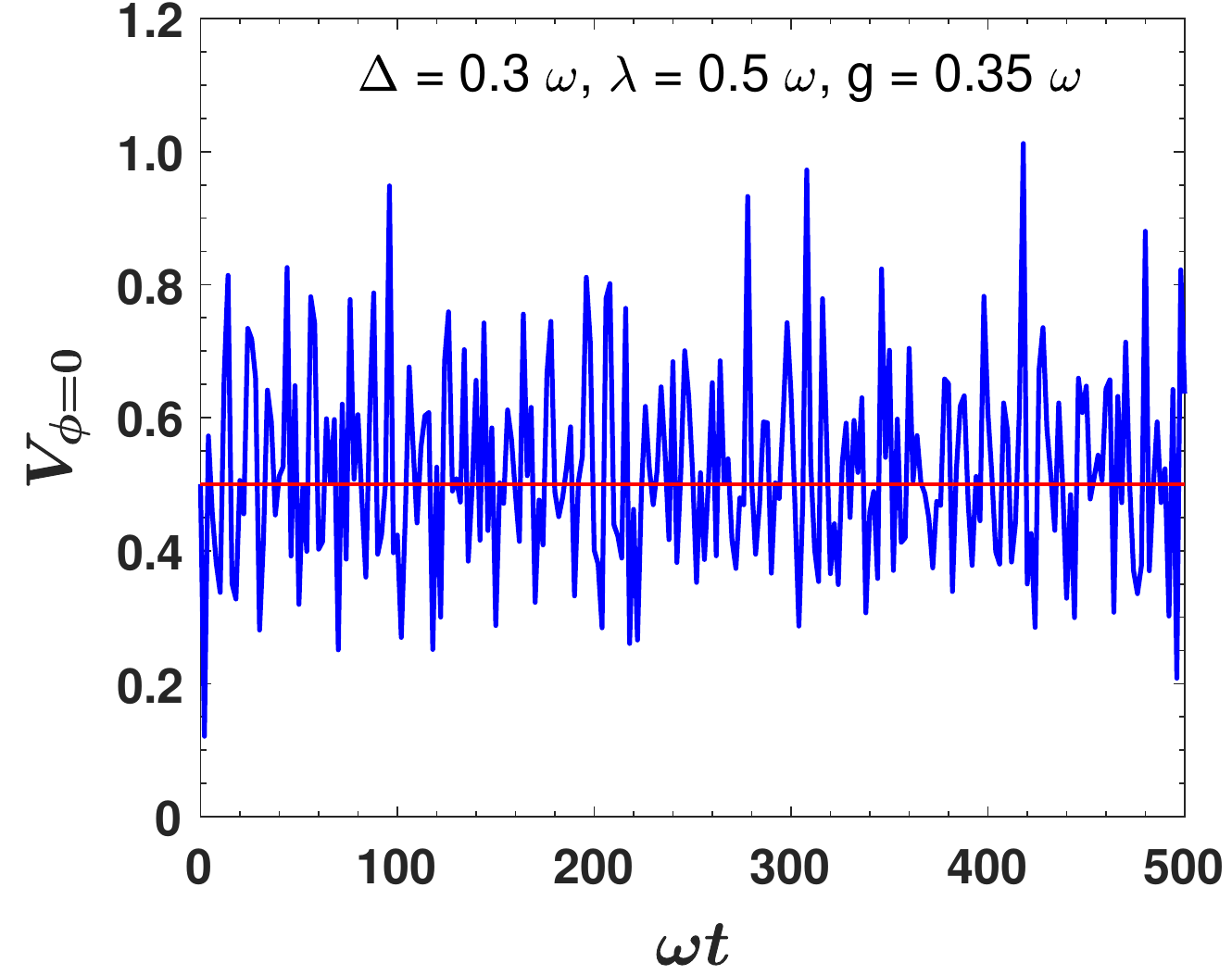}} 
		\captionsetup[subfigure]{labelformat=empty}
		\subfloat[$(\mathsf{c}_{1})$]{\includegraphics[width=4.5cm,height=4.5cm]{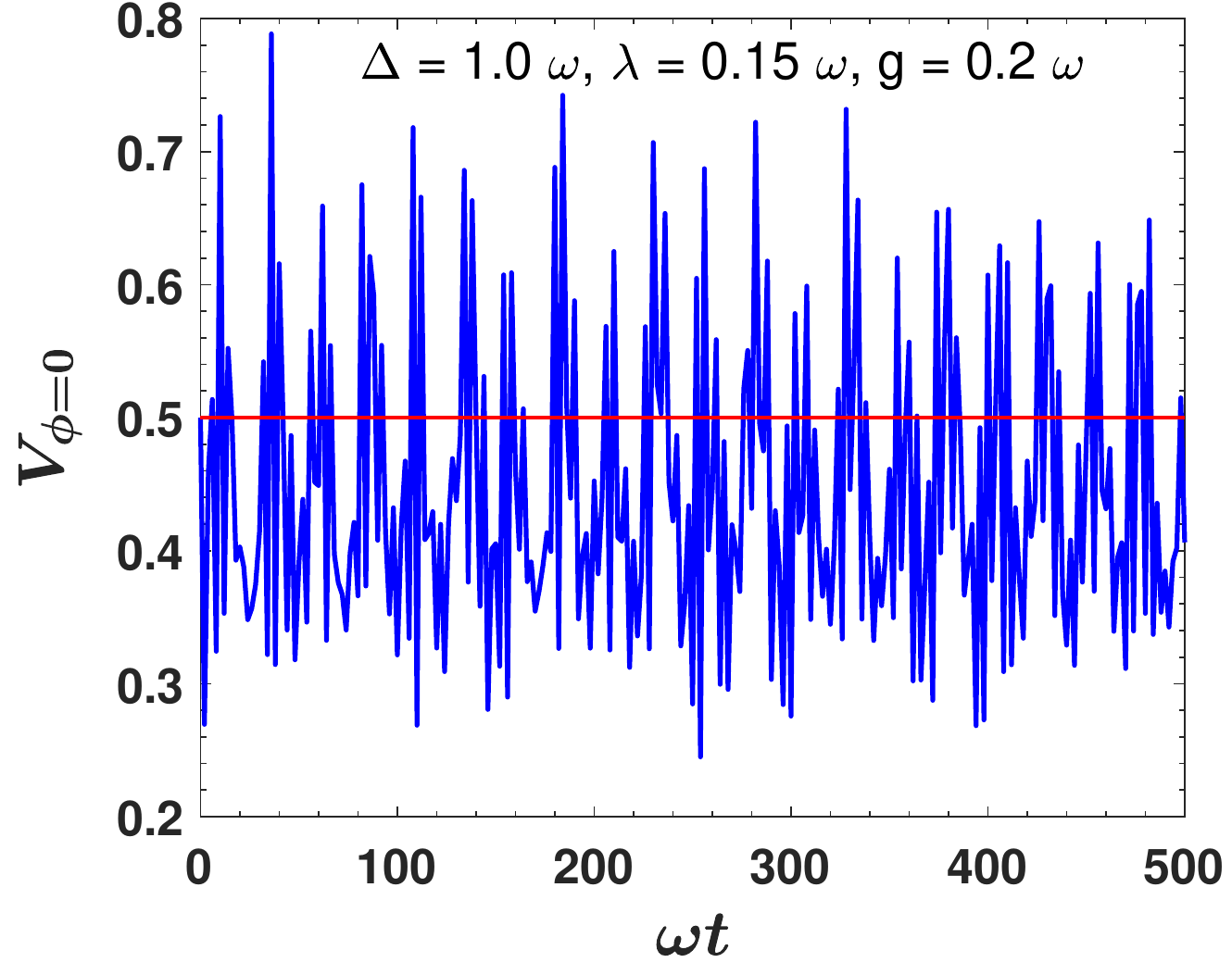}} 
		\captionsetup[subfigure]{labelformat=empty}
		\subfloat[$(\mathsf{c}_{2})$]{\includegraphics[width=3.5cm,height=3.5cm]{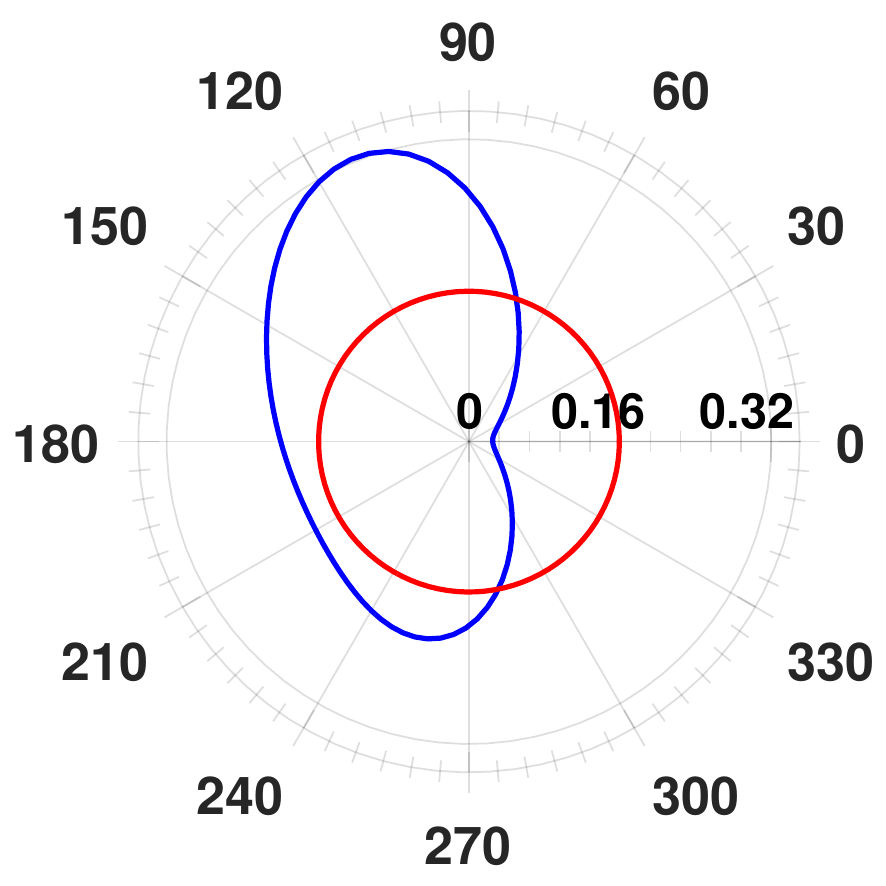}} 
		\captionsetup[subfigure]{labelformat=empty}
		\subfloat[$(\mathsf{c}_{3})$]{\includegraphics[width=3.5cm,height=3.5cm]{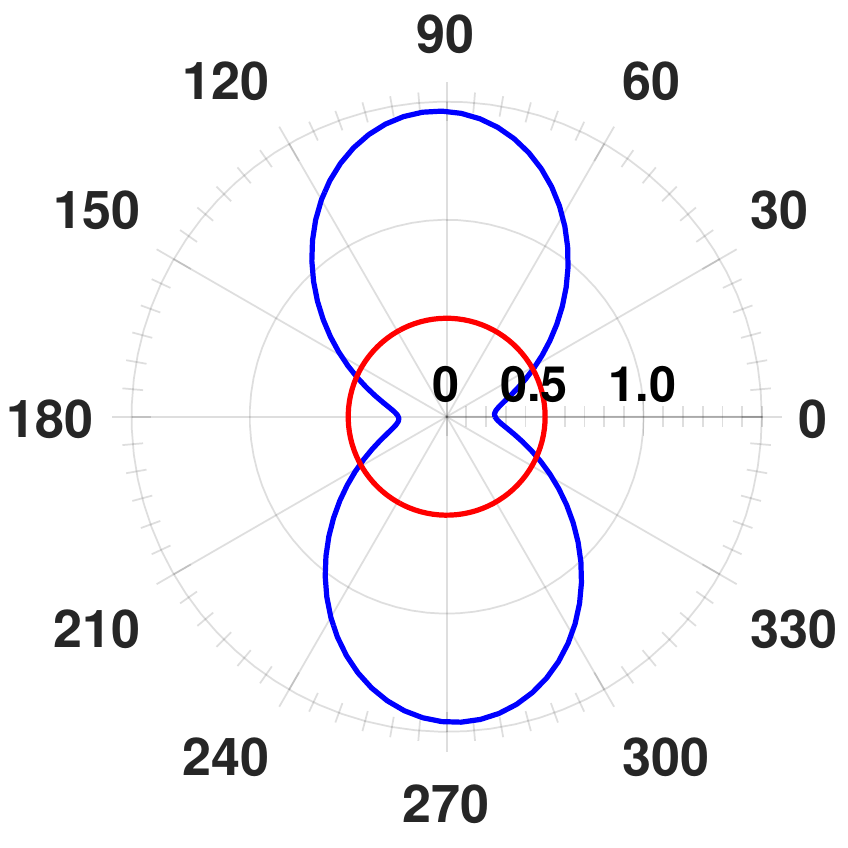}} 
		\captionsetup[subfigure]{labelformat=empty}
		\caption{(Color online). $(\mathsf{a}_{1})$ The time evolution of the quadrature variance $V_{\phi=0}$ (\ref{variance}) for the coupling constant $\lambda =\, 0.1 \,\omega$  at far away from the resonance $(\Delta = 0.3 \;\omega)$ in the absence of parametric nonlinear term $(g = 0)$. The horizontal red line represents the classical limit of the variance $V_{\phi} = 0.5$. The red circles in $(\mathsf{a}_{2})$, $(\mathsf{b}_{2})$ and $(\mathsf{c}_{2})$ indicate the polar phase density  of the $Q$-function (\ref{Q_polar}) for the vacuum state $\rho_{\mathcal{O}} = \ket {0}\bra {0}$ i.e. $\mathcal{Q}(\theta) = \frac{1}{2\pi}$ whereas the red circles in $(\mathsf{a}_{3})$, $(\mathsf{b}_{3})$ and $(\mathsf{c}_{3})$ describe the classical limit of the $V_{\phi} = 0.5$ at the scaled time $\omega t = 0$. The $\mathcal{Q}(\theta)$ in $(\mathsf{a}_{2})$ (blue) and the polar plot of $V_{\phi}$ in $(\mathsf{a}_{3})$ (blue) are denoted at $\omega t =220$.  The least value of the variance is observed at $\omega t=220$ equals to $0.4741$ $(\mathsf{a}_{1})$. The plot $(\mathsf{a}_{4})$ indicates that the squeezing is reduced with the increase of the coupling strength when $g =0$.
		The plot $(\mathsf{b}_{1})$  illustrates the same as $(\mathsf{a}_{1})$  in presence of the parametric nonlinear term $(g = 0.35 \, \omega)$. In this case the minimum value of $V_{\phi}$  equals to $0.0954$ which occurs at $\omega t = 264$. The $(\mathsf{b}_{2})$ and $(\mathsf{b}_{3})$ describe the same as $(\mathsf{a}_{2})$ and $(\mathsf{a}_{3})$ at $\omega t=264$ (blue) respectively for nonzero value of the parametric nonlinearity. $(\mathsf{b}_{4})$ The squeezing diminishes with the raising of the coupling strength even in presence of the parametric term.
		The plot $(\mathsf{c}_{1})$  shows the time evolution of the $V_{\phi}$ at the resonance $(\Delta = 1.0 \, \omega)$ for the coupling constant $\lambda = 0.15\, \omega$ and the parameter $g= 0.2\,\omega$. The $\mathcal{Q}(\theta)$  $(\mathsf{c}_{2})$ and polar plot of the $V_{\phi}$ $(\mathsf{c}_{3})$ (blue) reveal the squeezing effect evidently also in the case of resonance at $\omega t=254$.} 
		\label{SQ_PLT}
	\end{center}
\end{figure}
\par

 Another dynamical quantity that is useful in the study of squeezing is the polar phase density of the Husimi $Q$-function [\cite{RTM1993}] obtained via its radial integration on the phase space:
\bea
\mathcal{Q(\theta)}&=& \int\limits_{0}^{\infty} Q(\beta, \beta^{*})\, \lvert\beta\rvert \, d\lvert\beta\rvert, \quad \beta \,= \, \vert\beta\rvert \, \exp(i\theta),
\eea
which is a convenient tool for describing the splitting of the $Q$-function.
\bea
\mathcal{Q}(\theta) &=& \dfrac{1}{2}  |\mathcal{C}_{0}(t)|^{2} 
\left( \mathcal{I}^{(+)}_{0,0}(\theta) + \mathcal{I}^{(-)}_{0,0}(\theta) \right)
+ \mathrm{Re} \Big \lgroup \mathcal{C}_{0}(t)^{*} 
\sum_{n=1}^{\infty} \Big( \mathcal{A}_{n}(t) \left( \mathcal{I}^{(+)}_{0,n-1}(\theta) - \mathcal{I}^{(-)}_{0,n-1}(\theta) \right)\nn  \\
&+& \mathcal{B}_{n}(t) \left( \mathcal{I}^{(+)}_{0,n}(\theta) + \mathcal{I}^{(-)}_{0,n}(\theta) \right) \Big) + 
\sum_{n,m=1}^{\infty} \! \mathcal{A}_{n}(t)^{*} 
\mathcal{B}_{m}(t) \left( \mathcal{I}^{(+)}_{n-1,m}(\theta) - \mathcal{I}^{(-)}_{n-1,m}(\theta) \right) \Big \rgroup \nn \\
&+&\dfrac{1}{2}\! \sum_{n,m=1}^{\infty} \!\Big( \mathcal{A}_{n}(t)^{*} 
\mathcal{A}_{m}(t) \left( \mathcal{I}^{(+)}_{n-1,m-1}(\theta) + \mathcal{I}^{(-)}_{n-1,m-1}(\theta) \right)
+\mathcal{B}_{n}(t)^{*} \mathcal{B}_{m}(t) \left( \mathcal{I}^{(+)}_{n,m}(\theta) + \mathcal{I}^{(-)}_{n,m}(\theta) \right) \Big), \qquad
\label{Q_polar}
\eea
\bea
\mathcal{I}_{n,m}^{(\pm)}(\theta) \!\!\! &=& \!\!\! \frac{1}{\pi \tau_{_\theta}\sqrt{n!m!}} 
\left(\pm (\mu - \nu ) \frac{\eta}{\mu}\right)^{n+m} \exp\left(-\frac{\eta^{2}}{\mu}\left(\mu-\nu \right) \right) \sum_{k=0}^{n} \sum_{\ell=0}^{m}
\left( \pm \frac{\nu \sqrt{\mu \tau_{_{\theta}}}}{2\eta(\mu-\nu)}\right)^{k+\ell}\times \qquad \nn \\
&\times & k! \ell! \; \binom{n}{k}
\binom{m}{\ell}  \exp(i\theta(k-\ell)) \sum_{p=\lfloor \frac{k+1}{2} \rfloor}^{n} \sum_{q=\lfloor \frac{\ell}{2} \rfloor }^{m} \left( \frac{2}{ \nu \tau_{_\theta}}\right)^{p+q} \frac{(2p+2q+1-k-\ell)! }{p!q!}  \times \nn \\
&\times& \binom{p}{k-p} \binom{q}{\ell-q} \exp\left(-2i\theta (p-q)\right) 
\mathrm{H}_{k+\ell-2p-2q-2} \left(\pm \frac{\eta \cos \theta}{\sqrt{\mu \tau_{_\theta}}} \right),
\label{Qpol_abb}
\eea
where $\tau_{_\theta}= \mu + \nu \cos 2\theta$. The integrals employed to arrive at (\ref{Q_polar}) are listed below:
\bea
\int_{0}^{\infty} \!\!\!\!\!\!\!\!\!\! & & \!\!\!\!\! \mathrm{d}|\beta| \; |\beta| \exp\left(-|\beta|^{2}-\frac{\nu}{2\mu}(\beta^{2}+\beta^{*2}) 
\mp \frac{\eta}{\mu} (\beta + \beta^{*}) \right) \mathrm{H}_{n}\left(i \frac{\beta_{\pm}^{*}}{\sqrt{2 \mu \nu}}\right) 
\mathrm{H}_{m}\left(-i \frac{\beta_{\pm}}{\sqrt{2 \mu \nu}}\right) \nn \\
&=& \frac{\mu}{\tau_{_{\theta}}} (-1)^{m} \left( \pm i \frac{2  \eta(\mu-\nu)}{\sqrt{2\mu \nu}}\right)^{n+m}
 \sum_{k=0}^{n} \sum_{\ell=0}^{m} 
\left( \pm \frac{\nu \sqrt{\mu \tau_{_{\theta}}}}{2\eta(\mu-\nu)}\right)^{k+\ell}  k! \ell! \; \binom{n}{k}
\binom{m}{\ell} \times \nn \\
&\times& \exp(i\theta(k-\ell)) \sum_{p=\lfloor \frac{k+1}{2} \rfloor}^{n} \sum_{q=\lfloor \frac{\ell}{2} \rfloor }^{m} \left( \frac{2}{ \nu \tau_{_\theta}}\right)^{p+q} \frac{(2p+2q+1-k-\ell)! }{p!q!}  
\binom{p}{k-p} \binom{q}{\ell-q} \times \nn \\
&\times & \exp\left(-2i\theta (p-q)\right) \mathrm{H}_{k+\ell-2p-2q-2} \left(\pm \frac{\eta \cos \theta}{\sqrt{\mu \tau_{_\theta}}} \right).
\label{Qpol_int}
\eea
\section {The quadrature squeezing}
\setcounter{equation}{0}
\label{squeezing}
The quadrature operator is defined as
$X_{\phi}= \frac{1}{\sqrt{2}} (a \exp(-i\phi) + a^{\dagger} \exp(i\phi))$  where $\phi$ is a real phase [\cite{Barnett2002}].  The squeezing effect is characterized by the variance
\beq
V_{\phi} = \braket{X^{2}_{\phi}}-\braket{X_{\phi}}^{2}=  \mathrm{Re} \left( (\braket{a^{2}} -\braket{a}^{2}) \exp(-2i\phi) \right) + \braket{a^{\dagger} a} -  \; |\braket{a}|^{2} + \frac{1}{2}.
\label{variance}
\eeq
For the vacuum state as well as coherent states this variance is equal to $0.5$ which is called as the classical limit of the variance. The state of the field is said to be squeezed [\cite{Mandel1982}] if the corresponding variance is lesser than 0.5. The expectation values of the operators in the above variance can be conveniently computed via the $Q$-function through the following representation
\bea
\braket{a^{k}}=\int \mathrm{d}^{2}\beta \; \beta^{k} \; Q(\beta,\beta^{*}), \; \braket{a^{\dagger}a} = \braket{aa^{\dagger}}-1 = \int \mathrm{d}^{2}\beta \; |\beta|^{2} \; Q(\beta,\beta^{*})-1.
\eea
\bea
\braket{a^{k}} \!\! &=& \!\! \dfrac{1}{2}  |\mathcal{C}_{0}(t)|^{2} G^{(k,+)}_{0,0}
+ \frac{1}{2} \Big \lgroup \mathcal{C}_{0}(t)^{*} 
\sum_{n=1}^{\infty} \Big( \mathcal{A}_{n}(t) G^{(k,-)}_{0,n-1}
+ \mathcal{B}_{n}(t) G^{(k,+)}_{0,n} \Big) 
+ \mathcal{C}_{0}(t) 
\sum_{n=1}^{\infty} \Big( \mathcal{A}_{n}(t)^{*} G^{(k,-)}_{n-1,0} \nn \\
&+&
\mathcal{B}_{n}(t)^{*} G^{(k,+)}_{n,0} \Big) 
+
\sum_{n,m=1}^{\infty} \! \Big( \mathcal{A}_{n}(t)^{*} 
\mathcal{B}_{m}(t) G^{(k,-)}_{n-1,m} + 
\mathcal{A}_{n}(t) 
\mathcal{B}_{m}(t)^{*} G^{(k,-)}_{m,n-1}  \nn \\
&+&  \mathcal{A}_{n}(t)^{*} 
\mathcal{A}_{m}(t) G^{(k,+)}_{n-1,m-1}
+\mathcal{B}_{n}(t)^{*} \mathcal{B}_{m}(t) G^{(k,+)}_{n,m} \Big)  \Big \rgroup,
\label{exp_ak}
\eea
where the weight functions on the phase space read
\bea
G^{(k,\pm)}_{n,m}  \!\!\!\! &=&  \!\!\!\! \sqrt{n!m!}
\left(\frac{\nu}{2 \mu}\right)^{\!\! \frac{n+m}{2}} \sum_{\ell=0}^{k} ((-1)^{k-\ell}\mp 1) \binom{k}{\ell} (\eta (\mu - \nu))^{k-\ell} \left( \frac{\mu \nu}{2} \right)^{\ell} \sum_{p=0}^{n} \sum_{q=0}^{m} \frac{1}{p!q!} \binom{p}{n-p}
\binom{q}{m-q}
 \times \qquad \nn \\
&\times& \!\!\!\! \sum_{s=0}^{\min(2q+\ell-m,2p-n)} \left( \frac{2 \mu}{\nu} \right)^{s} s!
\binom{2q+\ell-m}{s} \binom{2p-n}{s} \mathrm{H}_{2q+\ell-m-s}(0) \mathrm{H}_{2p-n-s}(0).
\label{Q_weight}
\eea
To obtain the above expression, we make use of the following integrals
\bea
\int  \!\!\!\!\! \!\!\!\!\! & & \!\!\!\!\! \mathrm{d}^{2}\beta \; \beta^{k} \; \exp\left(-|\beta|^{2}-\frac{\nu}{2\mu}(\beta^{2}+\beta^{*2}) 
\mp \frac{\eta}{\mu} (\beta + \beta^{*}) \right) \mathrm{H}_{n}\left(i \frac{\beta_{\pm}^{*}}{\sqrt{2 \mu \nu}}\right) 
\mathrm{H}_{m}\left(-i \frac{\beta_{\pm}}{\sqrt{2 \mu \nu}}\right) \nn \\
&=&  \frac{\pi}{2}\, \mu \, i^{n} \, (-i)^{m} \, \sqrt{ n! m!} \left( \frac{2 \mu}{\nu}\right)^{\frac{n+m}{2}} 
\exp \left( \frac{\eta^{2}}{\mu} (\mu - \nu)\right) \left( G^{(k,+)}_{n,m} \pm G^{(k,-)}_{n,m} \right).
\eea
We also note that when $k=0$ in the above integrals, the finite summation series $G^{(k,\pm)}_{n,m}$ takes the values $G^{(0,+)}_{n,m} = 2 \; \delta_{n,m}$ and $G^{(0,-)}_{n,m} = 0$.
\bea
\braket{aa^{\dagger}} \!\! &=& \!\! \dfrac{1}{2}  |\mathcal{C}_{0}(t)|^{2} F^{(+)}_{0,0}
+ \frac{1}{2} \Big \lgroup \mathcal{C}_{0}(t)^{*} 
\sum_{n=1}^{\infty} \Big( \mathcal{A}_{n}(t) F^{(-)}_{0,n-1}
+ \mathcal{B}_{n}(t) F^{(+)}_{0,n} \Big) 
+ \mathcal{C}_{0}(t) 
\sum_{n=1}^{\infty} \Big( \mathcal{A}_{n}(t)^{*} F^{(-)}_{n-1,0} \nn \\
&+&
\mathcal{B}_{n}(t)^{*} F^{(+)}_{n,0} \Big) 
+
\sum_{n,m=1}^{\infty} \! \Big( \mathcal{A}_{n}(t)^{*} 
\mathcal{B}_{m}(t) F^{(-)}_{n-1,m} + 
\mathcal{A}_{n}(t) 
\mathcal{B}_{m}(t)^{*} F^{(-)}_{m,n-1}  \nn \\
&+&  \mathcal{A}_{n}(t)^{*} 
\mathcal{A}_{m}(t) F^{(+)}_{n-1,m-1}
+\mathcal{B}_{n}(t)^{*} \mathcal{B}_{m}(t) F^{(+)}_{n,m} \Big)  \Big \rgroup,
\label{Exp_aadgr}
\eea
\indent where
\bea
F^{(+)}_{n,m} &=& (-1)^{n} \frac{2}{\pi \mu \sqrt{n!m!}} \left(-\frac{\nu}{2\mu} \right)^{\frac{n+m}{2}}
\left( I_{n,m}^{(1,1)} + \pi \; \mu \; (\eta(\mu-\nu))^{2} \; n! \left(\frac{2\mu}{\nu}\right)^{n} \delta_{n,m}\right), \nn \\
F^{(-)}_{n,m} &=& (-1)^{n+1} \frac{2\eta(\mu-\nu)}{\pi \mu \sqrt{n!m!}}  \left(-\frac{\nu}{2\mu} \right)^{\frac{n+m}{2}}
\left(  I_{n,m}^{(1,0)} + I_{n,m}^{(0,1)}  \right)
\eea
\indent and
\bea
I_{n,m}^{(k,\ell)}  \equiv  \int \!\!\!\!\!\!\!\! & & \!\!\!\!\!\!\!\! \mathrm{d}^{2}\beta \; \beta^{k} \beta^{*\ell} \exp\left(-|\beta|^{2}-\frac{\nu}{2\mu}(\beta^{2}+\beta^{*2}) \right)  \mathrm{H}_{n}\left(i \frac{\beta^{*}}{\sqrt{2 \mu \nu}}\right) 
\mathrm{H}_{m}\left(-i \frac{\beta}{\sqrt{2 \mu \nu}}\right) \nn \\
&=& \!\!\!\! \pi \; \mu \; i^{n}(-i)^{m} n! m! \left( \frac{\mu \nu}{2}\right)^{\frac{k+\ell}{2}}
\sum_{p=0}^{n} \sum_{q=0}^{m} \frac{1}{p!q!} \binom{p}{n-p}
\binom{q}{m-q} \sum_{s=0}^{\min(\ell+2p-n,k+2q-m)} \left( \frac{2 \mu}{\nu} \right)^{s} s! \times \quad \nn \\
& & \times \binom{2p+\ell-n}{s} \binom{k+2q-m}{s} \mathrm{H}_{k+2q-m-s}(0) \mathrm{H}_{\ell+2p-n-s}(0).
\eea

\par

The enhancement of squeezing in the field mode is realized during the time evolution of the initial state of qubit-oscillator system in the presence of a parametric nonlinear term. In the strong coupling regime, the squeezing is noticed both at far away from resonance $(\Delta =0.3\,\omega)$ as well as at resonance $(\Delta=1.0\,\omega)$ (Fig. \ref{SQ_PLT}). The signature of the squeezing is observed when the variance $V_{\phi}$ of the quadrature variable, say at $\phi=0$ is rendered less than its classical value $1/2$. It is noticed that in the absence of parametric nonlinear term $(g=0)$, the least value of the variance $V_{\phi}$ reaches $0.4741$ at the scaled time $\omega t=220 $ for the coupling constant $\lambda =0.1\,\omega$  $(\mathsf{a}_{1})$. The  $\mathcal{Q}(\theta)$ and the polar plot of the variance $V_{\phi}$ (Fig. \ref{SQ_PLT} $(\mathsf{a}_{2})$ \& $(\mathsf{a}_{3})$ respectively) represent this quadrature squeezing more prominently. In presence of the parametric term, by comparing the Fig. \ref{SQ_PLT} $(\mathsf{a}_{1})$ and $(\mathsf{b}_{1})$ it is apparent that the enhancement of squeezing is happening for the parameter $g=0.35\,\omega$ with the identical coupling strength. It is important to note that the minimum value of the $V_{\phi}$ in this case is $0.0954$ at $\omega t=264$. This enhancement also reflects within the $\mathcal{Q}(\theta)$ and polar plot of the $V_{\phi}$ (Fig. \ref{SQ_PLT} $(\mathsf{b}_{2})$ \& $(\mathsf{b}_{3})$ respectively) at the same scaled time. 

\par
The squeezing in this qubit-oscillator model can be understood by suitably transforming the Rabi Hamiltonian (\ref{RPH}) under a unitary operation which allows the construction of the effective Hamiltonian [\cite{PHANGI2009}] in the dispersive limit i.e. $\lambda \ll \vert \Delta -\omega \rvert$ . The resulting effective Hamiltonian contains the two-photon terms $(a^{2}, {a^\dagger}^{2})$ that are responsible for the squeezing [\cite{CJS2017}]. However, the squeezing generated in the system is limited. Hence, the enhancement of squeezing to a notably large extent in the qubit-oscillator system can be achieved in the presence of a parametric nonlinear term which is obvious from the Fig. \ref{SQ_PLT} $(\mathsf{b}_{1})$. It is also noticed that the increase in the coupling strength in the strong coupling regime reduces the squeezing in the absence of nonlinear term as shown in the Fig. \ref{SQ_PLT} $(\mathsf{a}_{4})$.  This decrement in the squeezing happens due to the participation of other multiple photon terms in the effective Hamiltonian when we further increase the coupling strength. This higher order multi-photon terms become more significant compare to the two-photon terms and induce randomness in the phase relationship which considerably decreases the squeezing.

\par

The explicit presence of the parametric term in the system facilitates to overcome the aforesaid limitation and it can be understood as follows. The strength of the two-photon term in the Hamiltonian [\cite{PHANGI2009}] can be increased through the parameter $g$ and in this process the two-photon terms are allowed to dominate the other multi-photon processes. This leads to more squeezing generated in the system which is evident from the the Fig. \ref{SQ_PLT} $(\mathsf{b}_{1})$. This argument is also applicable in the case of resonance (Fig. \ref{SQ_PLT} $(\mathsf{c}_{1})$).

\section{Conclusion}
\label{sec_con}

We have studied an interacting qubit-oscillator bipartite system in the presence of a parametric nonlinear term by employing the generalized rotating wave approximation in the strong coupling domain. A comparison is outlined between the analytically obtained approximate energy spectrum with numerically computed spectrum of the full Hamiltonian to validate our approximation. For the initial state of the bipartite system, the time evolution of the reduced density matrix of the oscillator is obtained via the partial tracing over the qubit degree of freedom. On the oscillator phase space, its density matrix furnishes the Husimi $Q$-function with which we have derived the quadrature variance to realize the squeezing effect. It is observed that the squeezing gets reduced by increasing the coupling strength between the qubit and oscillator in the strong coupling limit. However, we have shown that the squeezing is enhanced significantly in the presence of a parametric term which corresponds to the two-photon process. This approach could be convincingly adopted to investigate the nonclassical features in the strongly interacting systems.

\section*{Acknowledgement}
We would like to thank M. Sanjay Kumar for his helpful comments and encouragement. One of us (PM) acknowledges the financial support from DST (India) through the INSPIRE Fellowship Programme.

\bibliographystyle{ieeetr}
\bibliography{References}
\end{document}